# Manganese Dioxide Decoration of Macroscopic Carbon Nanotube Fibers: From High-Performance Liquid-Based to All-Solid-State Supercapacitors

*Afshin Pendashteh,[a*] Evgeny Senokos,[a,b,c] Jesus Palma,[a] Marc Anderson,[a,d] Juan J. Vilatela,[c] and Rebeca Marcilla [a*]*

[a] Electrochemical Processes Unit, IMDEA Energy Institute, Parque Tecnológico de Móstoles, Avda. Ramón de la Sagra, 3, 28935 Móstoles, Madrid, Spain.

[b] E. T. S. de Ingenieros de Caminos, Universidad Politécnica de Madrid, 28040 Madrid, Spain.

[c] IMDEA Materials Institute, Tecnogetafe, Calle Eric Kandel, 2, 28906 Getafe, Madrid, Spain.

[d] Department of Civil and Environmental Engineering, University of Wisconsin, Madison, USA.

afshin.pendashteh@imdea.org; rebeca.marcilla@imdea.org

ABSTRACT

Supercapacitors capable of providing high voltage, energy and power density but yet light, low volume occupying, flexible and mechanically robust are highly interesting and demanded for portable applications. Herein, freestanding flexible hybrid electrodes based on $MnO_2$ nanoparticles grown on macroscopic carbon nanotube fibers (CNTf-$MnO_2$) were fabricated, without the need of any metallic current collector. The CNTf, a support with excellent electrical conductivity, mechanical stability, and appropriate pore structure, was homogeneously decorated with porous akhtenskite ε-$MnO_2$ nanoparticles produced via electrodeposition in an optimized organic-aqueous mixture. Electrochemical properties of these decorated fibers were evaluated in different electrolytes including a neutral aqueous solution and a pure 1-butyl-3-methylpyrrolidinium bis(trifluoromethylsulfonyl)imide ionic liquid ($PYR_{14}TFSI$). This comparison helps discriminate the various contributions to the total capacitance: (surface) Faradaic and non-Faradaic processes, improved wetting by aqueous electrolytes. Accordingly, symmetric supercapacitors with $PYR_{14}TFSI$ led to a high specific energy of 36 $Wh \cdot kg^{-1}_{MnO_2}$ (16 $Wh \cdot kg^{-1}$ including the weight of CNTf) and real specific power of 17 $kW \cdot kg^{-1}_{MnO_2}$ (7.5 $kW \cdot kg^{-1}$) at 3.0 V with excellent cycling stability. Moreover, flexible all solid-state supercapacitors were fabricated using $PYR_{14}TFSI$-based polymer electrolyte, exhibiting maximum energy density of 21 $Wh \cdot kg^{-1}$ and maximum power density of 8 $kW \cdot kg^{-1}$ normalized by total active material.

KEYWORDS: Carbon Nanotube Fiber; Manganese Dioxide; Ionic Liquid; Solid-state; Supercapacitor.

## 1. INTRODUCTION

Supercapacitors (SCs) are energy storage devices with high power density and excellent cycling stability, and hence have drawn extensive attention in the past two decades. To rival with devices based on chemical storage mechanisms, SCs need to be improved in terms of energy density. The simple equation, $E = \frac{1}{2} CV^2$, captures the two mechanisms to increase energy density (E) through increases in the capacitances of the active materials or augmenting the voltage window of the electrolyte.

In this regard, highly electrochemically stable aprotic ionic liquids (ILs) are very appealing for SC applications in comparison with aqueous or ordinary organic electrolytes. In addition to a broad electrochemical stability window, ILs have non-flammable properties, high chemical and thermal stability and negligible vapor pressure. They are safer electrolytes for SCs compared to organic ones currently used in commercial devices. Among different ILs, 1-butyl-3-methylpyrrolidinium bis(trifluoromethylsulfonyl)imide ($PYR_{14}TFSI$) is popular due to its favorable electrochemical window (> 5.5 V on platinum)[1] and good molar ionic conductivity (0.54 $S \cdot cm^2 \cdot mol^{-1}$) [2]. In fact, there are several examples of SCs operating up to 3.5 V using this IL in combination with activated carbons, graphene, CNTs, etc. as electrode materials [3, 4].

On the other hand, the capacitance of SCs has been improved by modifying and increasing specific surface area of carbon electrode materials, taking values of specific capacitance normalized by total active material mass from around 25 $F \cdot g^{-1}$ to a bit more than 100 $F \cdot g^{-1}$ [5]. These systems based mainly on electrostatic energy storage on porous carbon pores have an attractive simplicity, but still suffer from a relatively low energy density of an order of 5 $Wh \cdot kg^{-1}$ per cell, far from commercial batteries [5].

Unlike carbon-based materials, pseudo-capacitive materials such as metal oxides can accumulate charge through fast, surface-confined Faradaic reactions. Their high theoretical charge storage capacity makes them excellent candidates for high capacitance electrodes [6]. $MnO_2$ is probably the most promising of the metal oxides and has been extensively studied due to its low cost, high electrochemical activity, and environmentally benignity [7, 8]. Although most of these Faradaic reactions involve protons and hence require the use of aqueous electrolytes, $MnO_2$ has been recently shown to be active in pure organic electrolytes or mixtures with ILs through insertion/deinsertion of ions between its layered structure [9]. However, as with other semiconductors, its intrinsic poor electrical conductivity ($10^{-5}$ to $10^{-6}$ $S \cdot cm^{-1}$) limits its electrochemical performance. Overcoming this limitation is the main motivation behind exploring various $MnO_2$-based composites with carbonaceous materials [7, 10]. While the role of the carbons support is to act as current collector, the synthetic routes used often lead to hybrid materials with enhanced electrochemical properties due to synergistic combination of its components [11]. Among all carbonaceous materials, those that are self-standing percolated networks (e.g. CNT papers, Carbon fiber papers, etc.) are of particular interest. They can be directly used as conductive porous scaffolds, while also enabling their use in energy powering devices with augmented mechanical functions such as flexibility, stretch ability or even structural properties [12]. For instance, Zhang et al. electrodeposited $MnO_2$ on Ni plated filter paper as flexible substrates and employed them as positive electrodes in asymmetric SCs based on PVA-$Na_2SO_4$ gel electrolyte, achieving a volumetric energy density of 0.78 $mWh \cdot cm^{-3}$ [13]. In another study, Wang et al. electrodeposited $MnO_2$ on carbon cloth as positive electrode and integrated with $Fe_2O_3$/PPy nanoarray negative electrodes also on carbon cloth, achieving 0.22 $mWh \cdot cm^{-3}$ energy density and a power density of 165.6 $mW \cdot cm^{-3}$ [14]. Although these results show the success in fabricating freestanding flexible hybrid electrodes, these electrodes still suffer from low mechanical robustness, as substrates/current collectors used cannot withstand significant mechanical stresses. Furthermore, due to the large thickness of the carbon substrate in comparison with the metal oxide film and the fact that carbon does not participate or participates subordinately in charge storage, the volumetric energy and power densities are not satisfactory. More efficient development of $MnO_x$-based pseudocapacitors requires utilizing a carbon scaffold with high mechanical strength, large surface area, high conductivity, lightweight, and at a low volume fraction relative to the active materials, hence leading to higher volumetric energy and power densities. Macroscopic fibers of carbon nanotubes (CNTf) combine these properties with high-performance mechanical properties and are thus ideal materials for electrodes in SCs and other multifunctional and structural devices [15]. Indeed, we have demonstrated the excellent combination of mechanical and electrochemical properties of CNTf, particularly when combined with pure ionic liquids [16] and polymer electrolytes [17], leading for example to all-solid flexible electric double-layer capacitors stronger than copper. They are available as self-standing planar conductive porous electrodes onto which metal oxides and other materials can be readily deposited, which contrasts with the cumbersome preparation of conventional electrodes based on dispersions of carbon black, binders and active materials, that later need to be deposited onto metallic sheets of other current collectors and subjected to various annealing steps. And unlike metallic current collectors, CNTfs provide a substantial electrochemical double layer capacitance due to their large specific surface area (SSA) above 250 $m^2 \cdot g^{-1}$. These examples highlight new possibilities to interface $MnO_2$ with CNT fibers in the context of energy-storing textiles. In that respect the benefits of the materials and methods used in this work are the possibility to fabricate large-area planar electrodes with clearly define shape and size. This enables a direct comparison of properties with other materials, fundamental studies on the contribution of Faradaic and non-Faradaic processes, and ultimately the fabrication of devices with high energy density.

In the present work, we produce large-area planar electrodes based on arrays of macroscopic continuous CNT fibers onto which $MnO_2$ nanoparticles are grown through a simple electrodeposition technique. Electrochemical properties of these freestanding, binder- and additive-free hybrid electrodes have been evaluated in an aqueous electrolyte and in IL. In fact, to the best of our knowledge, this is the first report on $MnO_2$ decorated CNTf electrodes utilized to fabricate high voltage (3.0 V) supercapacitors with pure $PYR_{14}TFSI$ as the electrolyte. The comparison of properties in different electrolytes helps discriminate the various contributions to the total capacitance which includes surface pseudocapacitance, non-Faradaic processes, chemical capacitance and improved wetting by aqueous electrolytes. Moreover, to exploit the mechanical properties of these electrodes in a real device these CNTf-$MnO_2$ electrodes were integrated with $PYR_{14}TFSI$-based polymer electrolyte as solid self-standing and flexible supercapacitor, working at 3.0 V with high volumetric energy density and power density.

## 2. EXPERIMENTAL SECTION

**2.1. CNT Fibers Synthesis.** The pristine carbon nanotube fibers were spun directly from the gas phase during growth of the CNTs by chemical vapor deposition using ferrocene as Fe catalyst, butanol as the carbon source and thiophene as S promoter. The reaction was conducted under hydrogen atmosphere at 1250 °C. Precursor feed rate was set at 5 ml·h$^{-1}$ with a fiber winding rate of 3 m·min$^{-1}$. Detailed synthesis method can be ascertained in a previous report [16].

**2.2. $MnO_2$ Decorated CNT Fiber electrodes.** Electrodes were produced by first collecting CNT fibers onto a rotating spool and then densifying the unidirectional array with pure acetone.

This produces a thin porous fabric of consolidated CNT fibers. The array is then dried in a vacuum oven at 120 °C overnight. Such CNT fabric samples were carefully weighed with a microbalance (Mettler-Toledo, d = 1 μg), obtaining a mass loading in the range of 0.245 - 0.380 mg·cm$^{-2}$. A coupon of the CNT fiber (1×1 cm) fabric was employed as the working electrode using a platinum mesh and Ag/AgCl (ALS Co., RE-1B, 3 M NaCl) as counter and reference electrodes, respectively. CNT fibers fabrics were decorated by manganese dioxide nanoparticles through electrodeposition at a constant current density of 250 μA·cm$^{-2}$ for various time periods between 5 and 120 min with a multichannel Bio-Logic Science galvanostat/potentiostat (VMP3) station. It is worth to remark that due to the high electric conductivity of CNT fibers they were used directly as substrate without any current collector. The electrodeposition bath consisted of 0.05 M $Mn(NO_3)_2$ and 0.1 M $NaNO_3$ in a mixture of ethanol and distilled water (Mili-Q) with a volumetric ratio of 3.4:1 ($EtOH:H_2O$). The mixture composition was chosen among different combination ratios of water and ethanol in order to make sure full impregnation of CNT fibers (see Supplementary, Figure S1), complete dissolution of the solutes, and successful deposition of $MnO_2$ nanoparticles on the substrate. After electrodeposition, modified CNT fibers were thoroughly rinsed with acetone and dried under vacuum at 120 °C overnight. Samples were denoted as CNTf-Mn-$x$ where $x$ stands for the electrodeposition time period in min. After drying, the mass of $MnO_2$ deposit was estimated by weighing the sample, giving a mass loading between 0.035 and 0.880 mg·cm$^{-2}$ depending on the electrodeposition time. The $MnO_2$ to CNTf mass ratio was obtained as 0.01, 0.18, 0.37, 0.98, and 1.7 for samples prepared in 5, 10, 20, 60, and 120 min, respectively.

**2.3. Microstructural Characterization of the Samples.** X-ray Diffraction (XRD) patterns of the samples were recorded on a PANalytical Empyrean diffractometer equipped with Cu K$_\alpha$ radiation ($\lambda$ = 0.154 nm) at a generator voltage of 45 kV and an emission current of 40 mA. A FIB-FEGSEM Dual-beam microscope (Helios NanoLab 600i, FEI) was utilized to probe the morphology of the deposited samples on CNT fibers. For cross-section imaging of the samples, $MnO_2$ decorated CNT fibers were immersed in liquid nitrogen and were torn apart to make sure no bending occurs at the edges. Moreover, FEG S/TEM (Talos F200X, FEI) microscope was used to acquire TEM micrographs and elemental mapping. The samples were also examined by Raman spectroscopy using a JASCO NRS-5100 Laser Raman spectrometer equipped with green 532 nm laser. In addition, surface of the samples was analyzed by X-ray photoelectron spectroscopy (XPS) using a monochromatic Al $K_\alpha$ radiation with a voltage of 12 kV, an ion beam current of 6 mA, and energy resolution of 0.1 eV for high resolution spectra. $N_2$ adsorption–desorption measurements were conducted using a Quadrasorb instrument preceded by a thermal pretreatment of samples at 150 ℃ for 20 h.

**2.4. Electrochemical Measurements in Half-cell Configuration.** The electrochemical performance of the samples was investigated by cyclic voltammetry (CV) in three-electrode configuration in different media including aqueous (1 M $Na_2SO_4$) and ionic liquid (1-butyl-3-methylpyrrolidinium bis(trifluoromethylsulfonyl)imide, abbreviated as $PYR_{14}TFSI$, 99.5% pure, purchased from Solvionic, France). The samples were directly employed as self-standing working electrodes without using any current collectors. A Pt mesh was used as auxiliary electrode while Ag/AgCl and Ag spiral wire were employed as reference electrodes in aqueous and non-aqueous electrolytes, respectively. For non-aqueous media, the IL was taken from a glovebox ($O_2$ content < 1.0 ppm and $H_2O$ < 0.1 ppm) and Ar purging was continued above the solution during the experiment. The capacitances of the samples were calculated by integration of the area under CV curves using the following equation:

$$C_{sp} = \int i\, dV / mv\Delta V \qquad (1)$$

Where $C_{sp}$ is specific capacitance (e.g. capacitance normalized by mass of the active material) based on F.g$^{-1}$, $\int i\, dV$ is the integral under the curve, $m$ is the mass of active material (g), $v$ is the potential sweep rate (V·s$^{-1}$), and $\Delta V$ is the working potential window (V).

**2.5. Fabrication and Evaluation of Symmetric Supercapacitors.** Symmetric full devices were assembled by sandwiching two similar electrodes (e.g. two CNTf or two CNTf-Mn-60) with a Celgard separator in between and placing them in Swagelok cells in glovebox under inert atmosphere. Before mounting the electrodes and separator into the device they were soaked in PYR$_{14}$TFSI ionic liquid. In the case of all-solid-state supercapacitors, polymer electrolyte (PE) based on PYR$_{14}$TFSI ionic liquid (IL) (60% in weight) was prepared by sequent dissolving 5 mg of PVDF-co-HFP and 7.5 mg of IL in 30 ml of acetone and stirring for 5 h. Transparent self-standing PE membrane with a thickness of 100 μm was obtained by casting the solution using doctor blade technique and then drying at room temperature overnight and 5 h at 50°C under vacuum [17]. The self-standing PE membrane was then employed as both electrolyte and separator being placed in between two identical electrodes to fabricate flexible all-solid-state supercapacitors. Different electrochemical techniques including cyclic voltammetry (CV), galvanostatic charge-discharge (GCD) measurements, and electrochemical impedance spectroscopy (EIS) were used to assess the supercapacitor performance. EIS measurements were carried out at open circuit potential with a sinusoidal signal (amplitude of 10 mV) over a frequency range of 200 kHz to 100 mHz. Specific capacitance of the assembled supercapacitors was achieved from the slope of the discharge profiles according to the following equation:

$$C_{sp} = i / M \times slope = i\,\Delta t / M\Delta V \qquad (2)$$

Where $M$ is total mass of the active materials (CNTf and MnO$_2$ deposit) in both electrodes ($M = 2 \times m$). In addition, real energy density ($E_{real}$) and real power density ($P_{real}$) of the devices were estimated by integration of discharge profiles according to the following:

$$E_{real} = i \int V dt \tag{3}$$

$$P_{real} = {E_{real}}/{t_{discharge}} \tag{4}$$

3. **RESULTS AND DISCUSSION**

The CNT fibers were fabricated by a high-throughput direct CVD spinning method and collected as a unidirectional fabric by winding the material on itself and densified through capillary forces upon exposure to acetone (Figure S2). SEM micrograph shows that the fabric is composed of interconnected CNTs in the shape of fiber bundles (Figure S3) that overlapped on each other forming a porous network with high specific surface area ($S_{BET} \sim 270$ m$^2 \cdot$g$^{-1}$). The structure of CNT fibers (e.g. orientation and textural properties) was discussed in detail in our previous publications [16, 18, 19]. MnO$_2$ decorated CNT fibers were prepared through constant current electrodeposition, as it is schematically shown in Scheme 1.

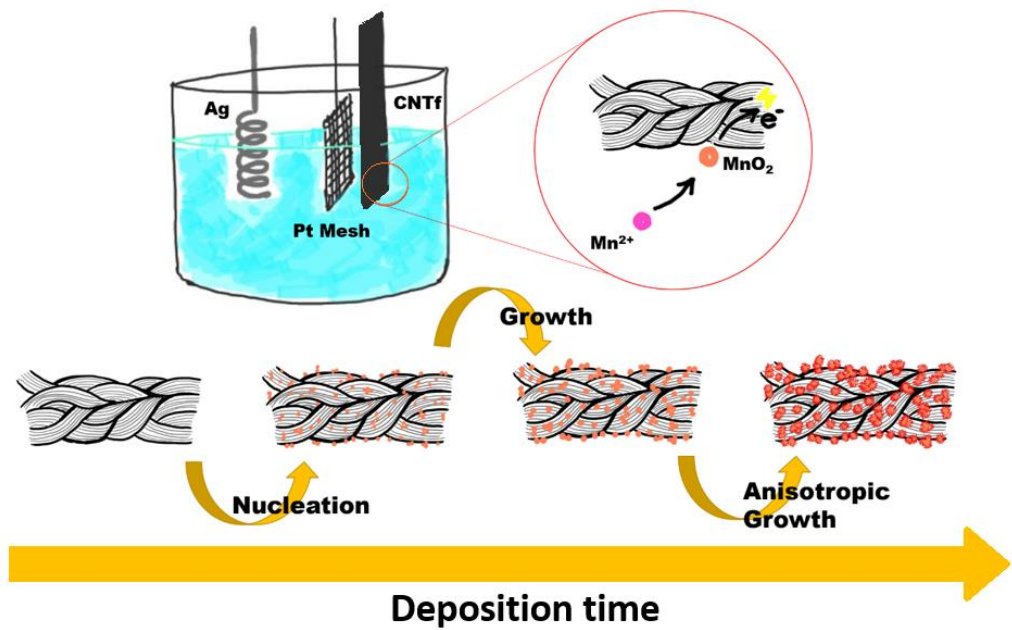

**Scheme 1.** Schematic illustration of MnO$_2$ NPs electrodeposition on CNT fibers and their anisotropic growth in prolonged deposition times.

Various samples were prepared by applying a current of 250 µA·cm$^{-2}$ for 5, 10, 20, 60 and 120 min. The prepared samples at various electrodeposition times were morphologically probed by FEG-SEM as shown in Figure 1 and S4. Figure S4a shows CNTf-Mn-5 sample in which MnO$_2$ nanoparticles can be distinguished clearly. The deposit consists of uniformly sphere-shaped particles with a diameter ranging from 80 to 125 nm which are well distributed all around the CNTf network (Figure S4b). Increasing the electrodeposition time into 10 min resulted in increased population of MnO$_2$ nanoparticles but with almost the same particle size range (Figure S4c and d). Further prolonging the electrodeposition, in CNTf-Mn-20 sample, all nanoparticles are clearly larger with an average diameter of 150 nm (Figure S4e and f). In Figure 1a, it can be seen that the surface of CNTf-Mn-60 sample is heavily covered with uniform sphere-shaped nanoparticles and the fibers cannot be distinguished in low magnification. At higher magnification (Figure 1b) it is clearly observed that the spheres are made of nanoflakes, forming porous spheres with an average diameter of 200 nm. In the sample prepared by electrodeposition for 120 min (Figure 1c), nanospheres enlarged further in size and their population was drastically increased as well. Interestingly, the fibers can still be noticed in higher magnification (Figure 1d), confirming that even at such a long electrodeposition time the fibers and their pores are still exposed and thus accessible to electrolyte ions. Figure 1e and f show typical cross-section images of the $MnO_2$ decorated CNTf sample (CNTf-Mn-60), showing that the thickness of the film is around 5 to 8 μm. It also demonstrates that not only the surface of CNT fibers is homogeneously covered with $MnO_2$ NPs but also they are well spread all along the thickness of the fiber film. However, it can be notified that the nanoparticles along the film thickness possess broader range of sizes with an alteration from sphere-shaped particles on surface to more irregular-shaped particles. This clearly confirms that the deposition solution could effectively impregnate CNTf, permitting nucleation of $MnO_2$ on the internal surface of the CNT fibers as well as in their pores. This results in a high mass loading of $MnO_2$ NPs homogeneously distributed in the CNTf scaffold in close contact and, thus benefiting from its use as both active material and current collector. In this fashion, the large surface area pseudocapacitive domains are electronically interconnected.

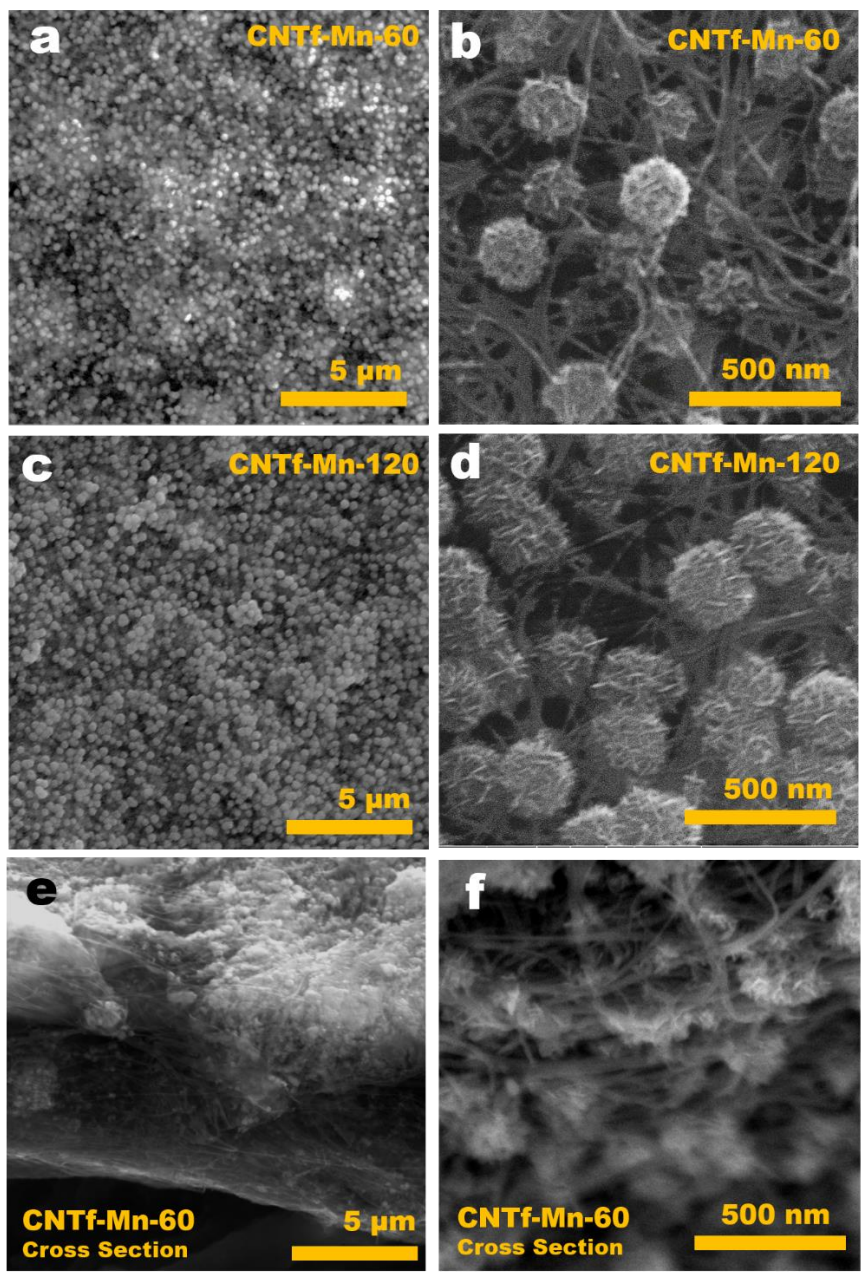

**Figure 1.** Morphological characterization of the samples: FEG-SEM micrographs of CNTf decorated by $MnO_2$ NPs through applying 250 µA·cm$^{-2}$ for a time period of (a, b) 60 and (c, d) 120 min. (e, f) cross section FEG-SEM image of decorated CNTf by $MnO_2$ NPs (CNTf-Mn-60), revealing good distribution of nanoparticles along the fiber thickness.

Growth of the $MnO_2$ nanoparticles on CNT fibers was further investigated through transmission electron microscopy and elemental mapping, as shown in Figure 2. At low magnification in Figure 2a, the image shows bundles of CNTs in the fiber, forming an interconnected network. The $MnO_2$ nanoparticles are in close contact with the CNTs, suggesting a strong interaction between the two phases. This strong interaction may be attributed to the possible formation of covalent bonds between $MnO_2$ particles and functional groups on the surface of CNT fibers through dehydration process; however, a clear understanding of the nature of these interactions needs further investigations. A closer view of $MnO_2$ domain in Figure 2b shows their structure comprised of fine nanoflakes aggregated in shape of dandelion flowers with an average diameter of 150 nm. These are in accordance with the observed morphology through SEM experiments. Distribution of elements was examined by STEM elemental mapping of decorated CNT fibers with $MnO_2$ NPs. High angle annular dark field scanning transmission electron microscographs (HAADF-STEM) and elemental maps are shown in Figure 2c-f. They confirm that the NPs observed by TEM are indeed $MnO_2$. The dandelion-shaped structure of the $MnO_2$ will result in higher accessible surface and consequently smaller diffusion path length for electrolyte ions which is of great importance in energy storage materials especially for high power applications [20-22].

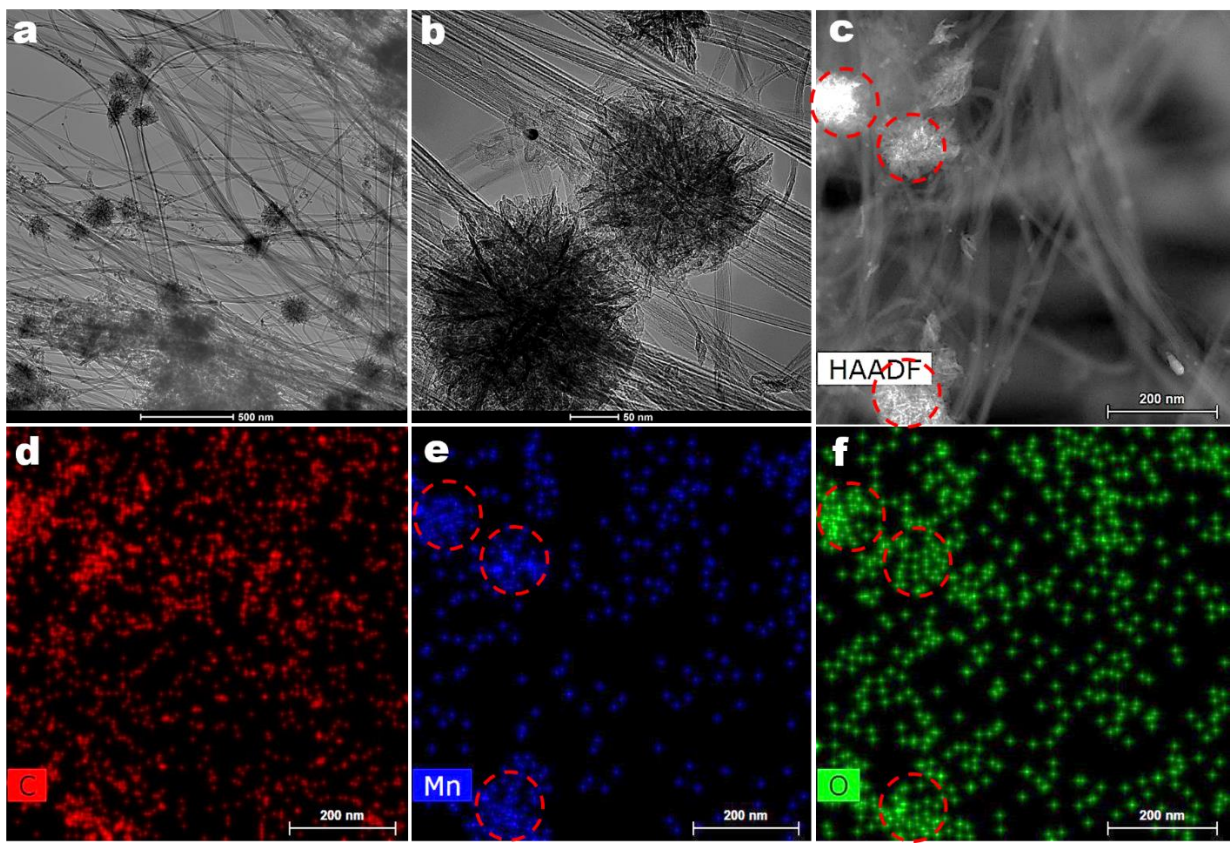

**Figure 2.** Morphological and elemental probing of the sample: (a, b) TEM micrographs of CNTf sample decorated by $MnO_2$ NPs (CNTf-$MnO_2$-60) at two different magnifications. (c) high angle annular dark field scanning transmission electron microscopy (HAADF-STEM) image of the sample and (d-f) corresponding EDX-STEM maps of various elements, depicting their distribution in the CNTf-$MnO_2$-60 sample.

X-ray diffraction (XRD) was utilized to examine the composition of the samples and the results for bare CNT fibers (CNTf) and decorated CNT fibers with $MnO_2$ are shown in Figure 3a. In comparison with the bare CNT fiber, two diffraction peaks clearly emerged after electrodeposition; one sharp peak at a 2theta of 37° and a broad, low intense peak at around 66° (marked with #). Relatively low signal-to-noise ratio and broad diffracted peaks suggest poor crystallinity of the oxide, however the pattern can be still well ascribed to hexagonal akhtenskite ε-MnO$_2$ (JCPDS no. 00-030-0820) and is in good agreement with previous reports [23-26]. Raman spectroscopy was also employed to further probe the formation of MnO$_2$ on CNT fibers, as shown in Figure 3b. Three sharp characteristic peaks of carbon nanotubes can be clearly seen in both samples. The G mode with E$_{2g}$ symmetry originates from in-plane bond-stretching motion of sp$^2$ C pairs and is centered at 1575 cm$^{-1}$, while D peak around 1345 cm$^{-1}$ corresponds to a mode of A$_{1g}$ symmetry associated with defects and sp$^3$ bonding [27, 28]. The peak at 2673 cm$^{-1}$ coincides with G' mode (also called 2D) and is a second-order two-phonon process [29]. Taking a more careful look at the spectra, it can be seen that I$_D$/I$_G$ intensity ratio increased from 0.46 in bare CNT fibers to 0.60 in CNTf-Mn-60 sample. This is attributed to the formation of defects on the CNTs formed when they are interfaced with the MnO$_2$ nuclei and which give an indication of the strong interaction between the two phases. In the sample decorated with MnO$_2$ a peak at ~645 cm$^{-1}$ attributed to Mn-O stretching vibration of MnO$_2$ is also observed [25, 30, 31].

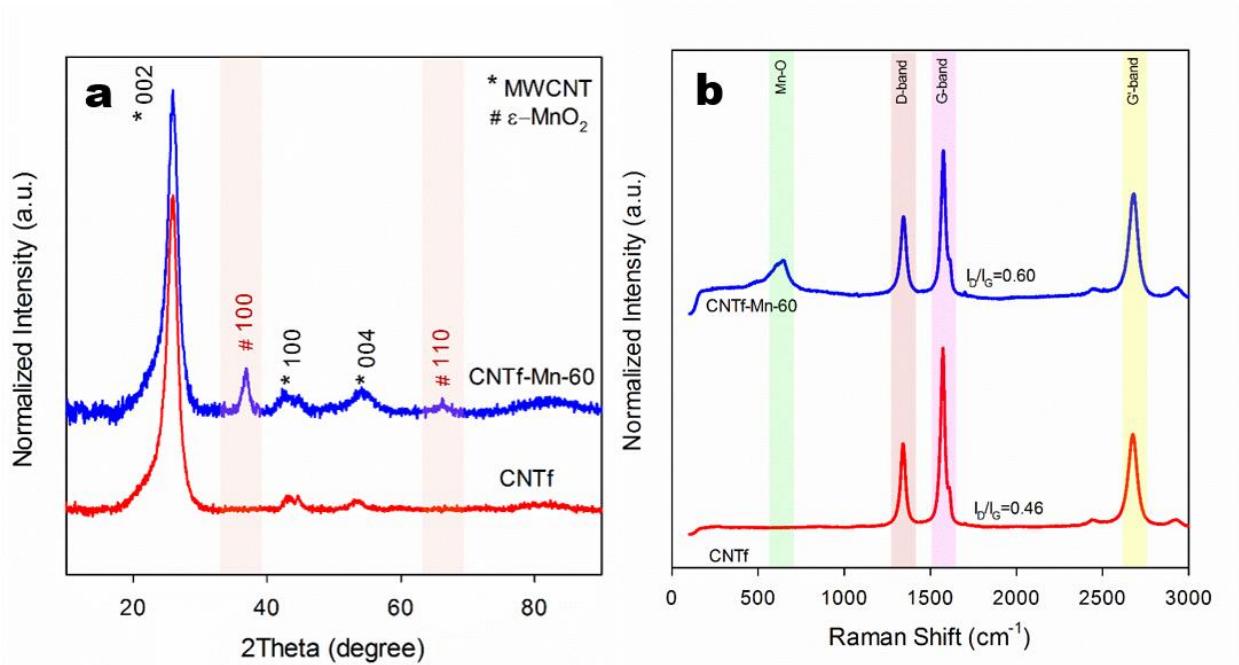

**Figure 3.** Compositional characterization of the samples: (a) XRD patterns of the pristine CNTf and MnO$_2$ deposited CNTf (CNTf-Mn-60); (b) Raman spectra of the pristine CNTf and modified sample with MnO$_2$ NPs during 60 min.

In order to explore elemental composition and chemical oxidation states of manganese oxide nanoparticles on CNTf bundles, X-ray photoelectron spectroscopy was carried out. The overall survey spectrum (Figure S5) demonstrates the existence of manganese, oxygen, and carbon in the CNTf decorated with MnO$_2$ NPs (CNTf-Mn-60). Deconvolution of high-resolution C 1s spectrum (Figure 4a) reveals one main sharp (~ 284.6 eV) and three less intense, broad peaks centered at 285, 286.4, and 290.1 eV, corresponded to C=C (graphitic sp$^2$ carbon atoms), C−C (sp$^3$ carbons), C−O (carbon in epoxy/ether groups), and O=C−O respectively [32-34]. As it is seen in Figure 4b, two characteristic spin-orbit doublets (Mn 2p$_{3/2}$ and 2p$_{1/2}$ centered at 642.0 and 653.7 eV, respectively) with a spin-energy splitting of 11.7 eV in high-resolution spectrum of Mn 2p appeared, revealing a predominant +4 oxidation state of the manganese oxide nanoparticles. The observed peak positions and splitting is in excellent agreement with the Mn 2p XPS data reported in MnO$_2$ [10, 35, 36]. Figure 4c depicts core level of O 1s, consisting of four components; one centered at 529.8 eV which can be attributed to Mn−O−Mn and the others at 530.9, 531.7, and 533 eV, respectively originated from Mn−OH, and C−O and C=O originating from the groups on CNT fibers [37].

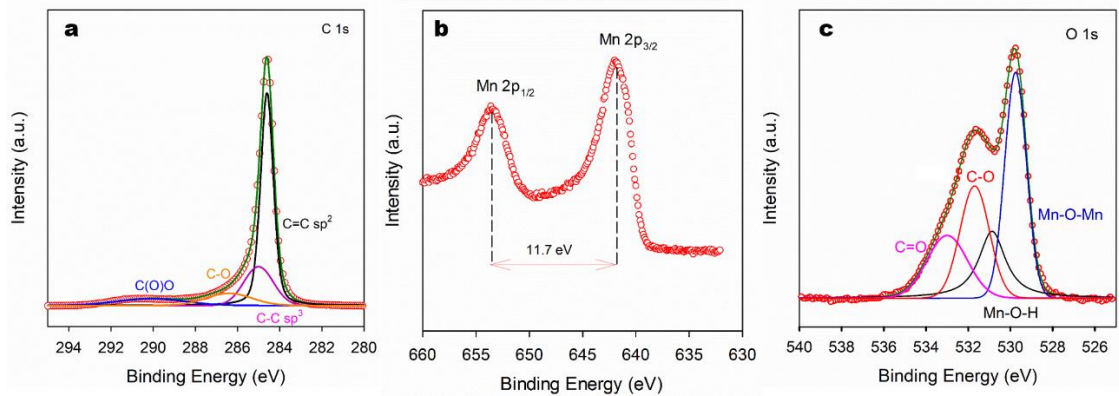

**Figure 4.** Surface characterization of MnO$_2$ decorated CNTf sample (CNTf-Mn-60): high resolution XPS core level of (a) C 1s, (b) Mn 2p, and (c) O 1s.

The modified CNT fibers with different loadings of MnO$_2$ were directly employed as self-standing electrodes without any current collector and their electrochemical behavior was examined in 3-electrode electrochemical cells first in neutral aqueous electrolyte (1 M Na$_2$SO$_4$). Figure 5a shows cyclic voltammograms at a scan rate of 50 mV·s$^{-1}$ of CNTf/MnO$_2$ hybrid samples with different metal oxide mass fraction. The graph also includes data for bare CNTf for reference, which is very low because of its hydrophobicity and the corresponding lack of aqueous electrolyte infiltration. As it can be seen, a very well rectangular shape CV signal was obtained for all hybrid samples, as in ideal supercapacitors [38]. Moreover, the area under the CV curve is significantly increased in the case of decorated CNTf samples with MnO$_2$ NPs in comparison with the pristine CNTf. This increase confirms the contribution of pseudocapacitance of MnO$_2$ NPs. The specific capacitance was calculated by integration of CV curves and normalized either by the mass of MnO$_2$ NPs or by the mass of whole electrode (CNTf + MnO$_2$ deposit) (red and blue curves in Figure 5b, respectively). As it is seen, the total specific capacitance increased by increasing the deposition time, reaching a maximum capacitance of 87 F·g$^{-1}_{\text{total}}$ in CNTf-Mn-120 sample. It is worthy to mention that the specific capacitance values in literature for $MnO_2$ ordinary electrodes usually do not include the weight of current collectors and necessary additives such as conducting agent and binder, resulting in very high specific capacitances which cannot be achieved in real devices after cell packing. This is not the case in these binder- and additive-free self-standing electrodes that do not need current collector. Considering only the mass of $MnO_2$, the highest specific capacitance of 310 $F·g^{-1}_{MnO_2}$ was obtained for the CNTf-Mn-5 sample, whereas increasing the deposition time led to gradual decrease of these values to 145 $F·g^{-1}$ for CNTf-Mn-120 sample. This is attributed to the surface-confined nature of the pseudocapacitance reactions and the fact that longer deposition times reduce the density of active sites per unit mass. But interestingly, although longer electrodeposition times reduce the specific capacitance of the NPs, the total specific capacitance keeps rising. This is a clear indication that the presence of metal oxide NPs also enables the infiltration of aqueous electrolyte into the otherwise hydrophobic CNTf, thus increasing the contribution from the EDLC of CNTf local areas free of metal oxide.

Figure 5c shows the rate capability of different CNTf-Mn-*x* samples at various scan rates. In longer deposition times, the samples show inferior rate capability due to a decreased electrical conductivity of the electrodes as the semiconducting metal oxide particles increase in size and so does electron flow path from the core to the conducting CNTf network. However, the specific capacitance values for longer electrodeposition times are superior at all measured scan rates. As it is seen, values for the samples prepared at 60 and 120 min are very close to each other, so the CNTf-Mn-60 sample was chosen as the optimum and all next experiments were conducted on this sample. It is worthy to mention that the $MnO_2$ mass fraction in the case of CNTf-Mn-60 sample is around 0.96, revealing a high mass loading of $MnO_2$ with respect to CNTf.

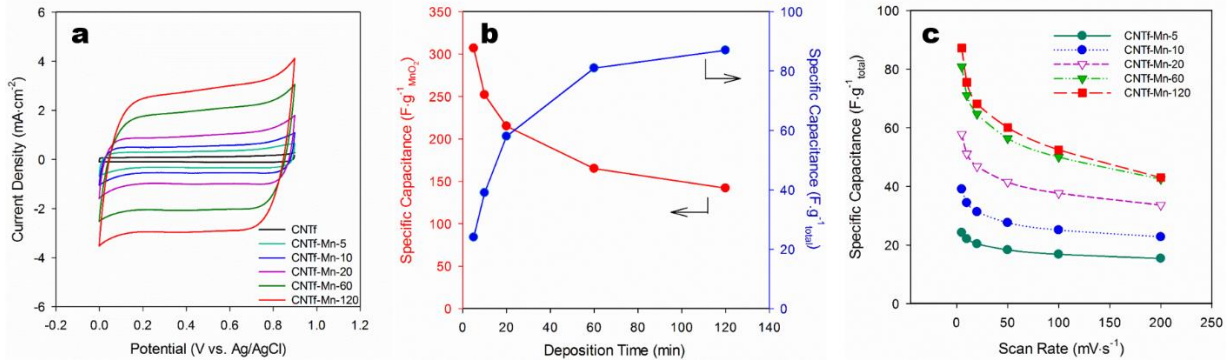

**Figure 5.** Electrochemical characterization of various samples in neutral aqueous medium: (a) cyclic voltammograms of the pristine and decorated CNTf samples (CNTf-MnO$_2$-$x$) at 50 mV·s$^{-1}$ in 1 M Na$_2$SO$_4$ electrolyte; (b) specific capacitance evolution in various deposition times based on the total electrode mass and MnO$_2$ deposit at a scan rate of 5 mV·s$^{-1}$; and (c) rate capability of the samples at various scan rates ranging from 5 to 200 mV·s$^{-1}$.

In an attempt to boost the operating voltage of the final device, electrochemical behavior of the bare and modified CNTf was examined in PYR$_{14}$TFSI ionic liquid, which provides a wide electrochemical stability window. In the case of pure CNTf (Figure S6), cyclic voltammetry measurements in 3-electrode configuration at different scan rates resulted in well-defined butterfly-shaped voltammograms. This behavior is typical of quasi-metallic and highly graphitic carbons such as SWCNTs [39-41], and arises from their quantum (chemical) capacitance [16, 42]. Figure 6a shows CV curves for CNTf-Mn-60 sample in pure PYR$_{14}$TFSI electrolyte at various scan rates. Interestingly, the CVs are more rectangular in shape loosing most traces of the butterfly-shape observed in pristine CNTf (Figure S6). This is mainly attributed to the high contribution of MnO$_2$ pseudocapacitance that thus makes the CNTf quantum capacitance negligible. In this respect, whereas the EDLC and quantum capacitance can be visualized as two elements in series, the pseudocapacitive contribution from the metal oxide would be in parallel with the total capacitance of the CNTf and thus dominates in the CV measurement. Nevertheless, we anticipate that the presence of MnO$_2$ NPs in close contact with the CNTf could have an effect on its quantum capacitance of the type observed with dopants,[43-45] charged impurities or functional groups.

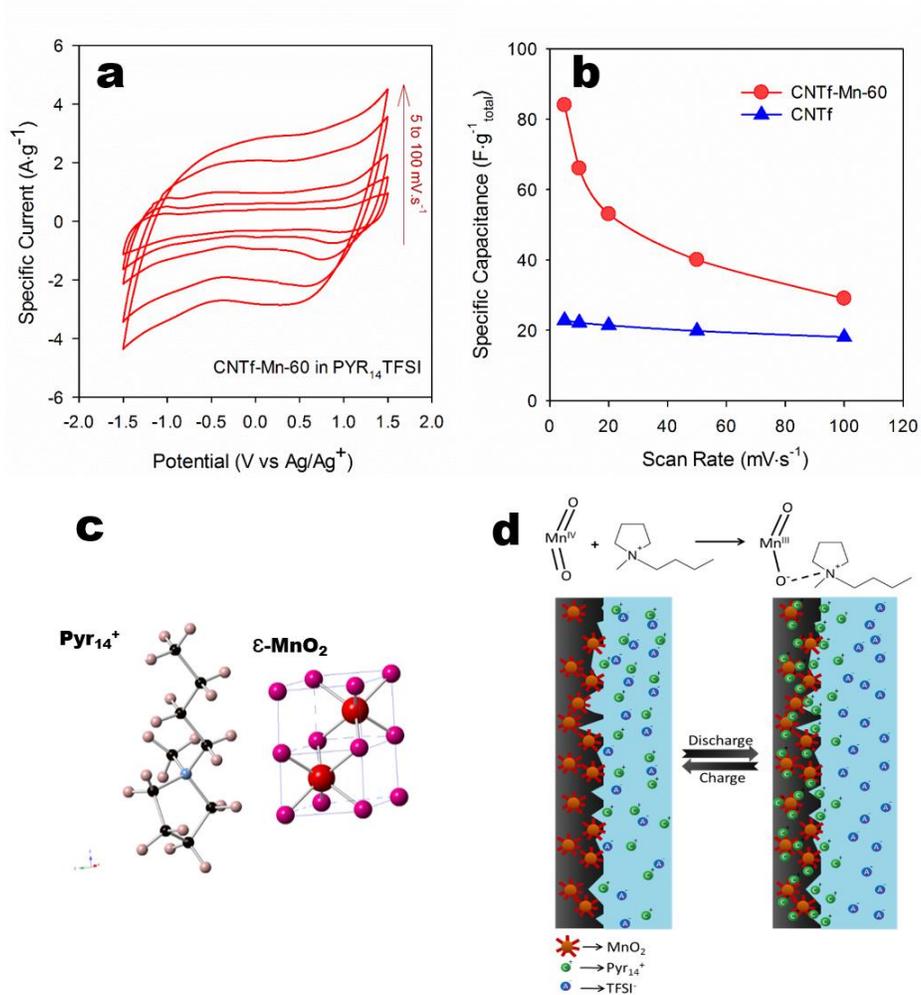

**Figure 6.** Electrochemical characterization of the samples in pure PYR$_{14}$TFSI ionic liquid: (a) cyclic voltammograms of CNTf-MnO$_2$-60 at various scan rates and (b) rate capability of bare CNTf and CNTf-MnO$_2$-60 sample; (c) Comparison of 1-butyl-1-methylpyrrolidinium cation and the unit cell of akhtenskite ε-MnO$_2$; and (d) the schematic illustration of charge storage mechanism of MnO$_2$ in PYR$_{14}$TFSI ionic liquid.

Figure 6b depicts evolution of specific capacitance for CNTf-Mn-60 sample over different scan rates. Interestingly, at a low scan rate of 5 mV·s$^{-1}$ a specific capacitance as high as 84 F·g$^{-1}$ was obtained (around 4 times greater than the pristine CNT fibers) and comparable to the value achieved in aqueous electrolyte. This does not usually happen in microporous carbon samples whose capacitances in IL are usually lower than in aqueous electrolyte since ions of IL are usually larger than the small pores. The comparable capacitance of decorated CNTf in aqueous and IL electrolyte arise from the open porosity of the hybrid sample, containing channels for diffusion of the relatively large IL ions. The sample could retain a capacitance of around 30 F·g$^{-1}$ even at a high scan rate of 100 mV·s$^{-1}$. These observations demonstrate excellent contribution of MnO$_2$ nanoparticles in charge storage phenomenon and suggests CNTf-Mn-60 sample as a promising electrode material in pure aprotic IL-based supercapacitors.

The mechanisms for charge storage in MnO$_2$ in aqueous electrolytes and mixtures of aqueous electrolytes with organic solvents, are reasonably well understood [43]. But to our knowledge, there are no reports using aprotic ionic liquids and, it is thus of interest to analyze the possible charge storage mechanisms operating in these conditions. The SSA of the CNT fibers (270 m$^2$·g$^{-1}$) is much greater than the SSA of the composite sample (Figure S7), hence the increase in capacitance of the hybrid material in IL cannot be ascribed to a large surface area. MnO$_2$ acts as a typical pseudocapacitive metal oxide storing energy through Mn$^{4+}$ ⇌ Mn$^{3+}$ reversible redox reaction. The general mechanism of manganese oxide during change of metal oxidation state is described by following redox reactions:

$$(MnO_2)_{surface} + C^+ + e^- \rightleftarrows (MnOOC)_{surface}$$

$$MnO_2 + C^+ + e^- \rightleftarrows MnOOC,$$

where $C^+$ is the electrolyte cation. The first reaction represents faradaic process on the surface of $MnO_2$, whilst the second corresponds to intercalation of the cations in the bulk of the oxide [43, 44]. Ionic intercalation can be ruled out on account of the much larger $Pyr_{14}^+$ cation size relative to the octahedral interstitials in $\varepsilon$-$MnO_2$ (Figure 6c). Linear relationship of the average current against square root of scan rate ($I_{avg.}$ vs. $v^{1/2}$) confirms mass transport controlled of the reaction, probably due to diffusion of the IL cation to the surface of $MnO_2$ (Figure S8) [45]. However, it should be also mentioned that the charge storage mechanism for $MnO_2$ in ILs might be affected by the crystal water which usually exists in $MnO_2$ lattice [46-48]. Further investigations are ongoing in order to shed more light on $MnO_2$ charge storage mechanism in ILs.

Galvanostatic charge-discharge (GCD) experiments were also conducted at various specific currents (Figure 7a) to further investigate the charge storage behavior of $MnO_2$ decorated CNTf samples. As it is seen, GCD profiles are almost triangular shape with slight deviations originated from pseudocapacitance of $MnO_2$ NPs. Moreover, the symmetric line shape of the profiles demonstrates good reversibility of the reaction as well as excellent coulombic ($\mu=Q^+/Q^-$) efficiency of around 97% at 1 A·g$^{-1}$ (Figure 7a). Electrochemical Impedance Spectroscopy (EIS) technique was also utilized to probe the electrochemical properties of CNTf-Mn-60 sample. Figure 7b represents Nyquist plot of the sample at open circuit potential (OCP). Accordingly, an ESR value (combination of electrolyte and material intrinsic resistances) of around 53 $\Omega$·cm$^{-2}$ was achieved for CNTf-Mn-60 sample which is greater than that of pristine CNTf (12 $\Omega$·cm$^{-2}$). This is originated from the intrinsically low electrical conductivity of $MnO_2$ NPs, leading to a low capacity retention compared with bare CNTf, as observed in Figure 6b. At low frequencies, an almost vertical line appeared which is an indication of typical capacitive behavior. The overall impedance is inversely proportional to the capacitance and that of CNTf-Mn-60 sample is much smaller than the pure

CNTf which demonstrates competence of the MnO₂ decorated sample for supercapacitor applications. Overall, no indication of diffusion limitations or a large CNT/MnO₂ charge transfer resistance were observed from the impedance data.

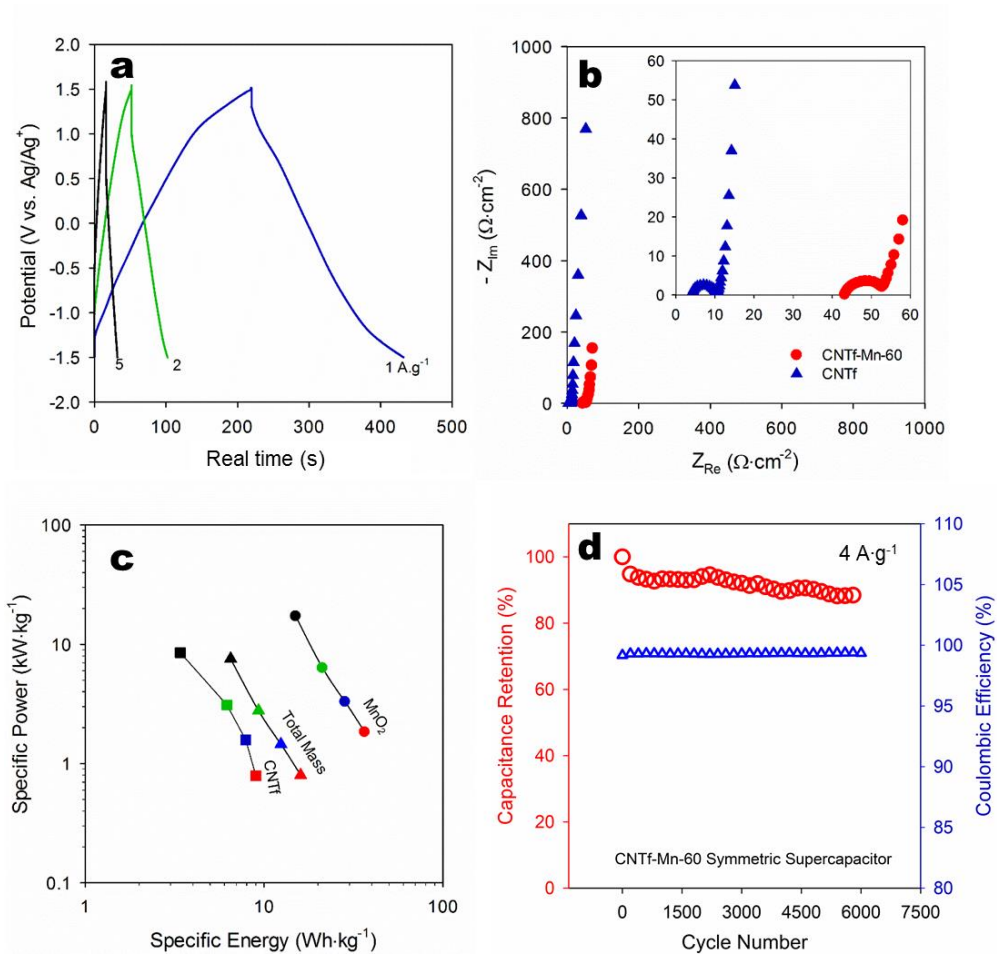

**Figure 7.** Electrochemical characterization of the samples in pure PYR₁₄TFSI ionic liquid: (a) galvanostatic charge-discharge profiles of CNTf-MnO₂-60 sample at various current densities; and (b) Nyquist plot of bare CNTf and CNTf-MnO₂-60 samples. Performance evaluation of CNTf-MnO₂-60 symmetric supercapacitors in pure PYR₁₄TFSI ionic liquid: (c) Ragone plot of the CNTf-MnO₂-60 symmetric supercapacitors based on total electrode mass and MnO₂ deposit as well as the one for CNTf symmetric supercapacitors; and (d) cycling stability and Coulombic efficiency of CNTf-MnO$_2$-60// CNTf-MnO$_2$-60 symmetric cells over 6000 cycles at 4 A·g$^{-1}$.

To demonstrate the applicability of MnO$_2$ decorated CNTf samples in practice, full symmetric devices were assembled and examined in PYR$_{14}$TFSI electrolyte. GCD profiles of the symmetric supercapacitors in 3.0 V at various specific currents were shown in Figure S9 b and c. According to calculations a specific capacitance of around 14 F·g$^{-1}$ was achieved at 0.5 A·g$^{-1}$ based on total mass of electrodes (2 × (CNTf + MnO$_2$ deposit)), equaling to a high specific capacitance of 56 F·g$^{-1}$ for a single electrode. Corresponding Ragone plot is shown in Figure 7c, revealing a high specific energy of 36 Wh·kg$^{-1}_{MnO_2}$ and a maximum specific power of ~17 W·kg$^{-1}_{MnO_2}$ (16 Wh·kg$^{-1}$ and 7.5 kW·kg$^{-1}$ according to total mass). Moreover, as it is seen the obtained values are much higher (c.a. twice greater) than the symmetric supercapacitor based on pristine CNTf electrodes, clearly demonstrates superior performance of the supercapacitor based on MnO$_2$ decorated CNTf electrodes. EIS measurements on symmetric device resulted in a complex plan plot, presented in Figure S9d. As is seen, the curve shows an almost vertical straight line for typical supercapacitors with a small ESR value, representing promising impedance properties of the cell. Cycling stability of CNTf-MnO$_2$-60//CNTf-MnO$_2$-60 symmetric cells has been examined with consecutive charge and discharge cycles at different current densities. Figure S9e shows cycling profile of the device at a current density of 2 A·g$^{-1}$ over 1250 cycles. As is seen, the capacitance increases during first 20 cycles and later decreases slightly, showing excellent cycling stability and Coulombic efficiency with almost no capacitance fading. Gradual capacitance increase specially at lower current densities has been previously seen and reported for metal oxide electrodes and their composites with carbon [49-51]. This increase in capacitance is ascribed to the electrochemical or structural activation caused by pore opening during charge-discharge cycles. Moreover, cycling performance of the device was further evaluated at a current density as high as 4 A·g$^{-1}$ over 6000 cycles and the outcome has been displayed in Figure 7d. It can be seen that the device could retain around 90% of its initial capacitance after 6000 cycles, revealing promising performance stability of the CNTf-MnO$_2$-60//CNTf-MnO$_2$-60 symmetric supercapacitors at high current densities. Furthermore, SEM imaging of the sample after cycling (Figure S10) demonstrated that the morphology of the MnO$_2$ NPs remained almost unchanged during consecutive charge-discharge cycles, revealing excellent stability of the sample.

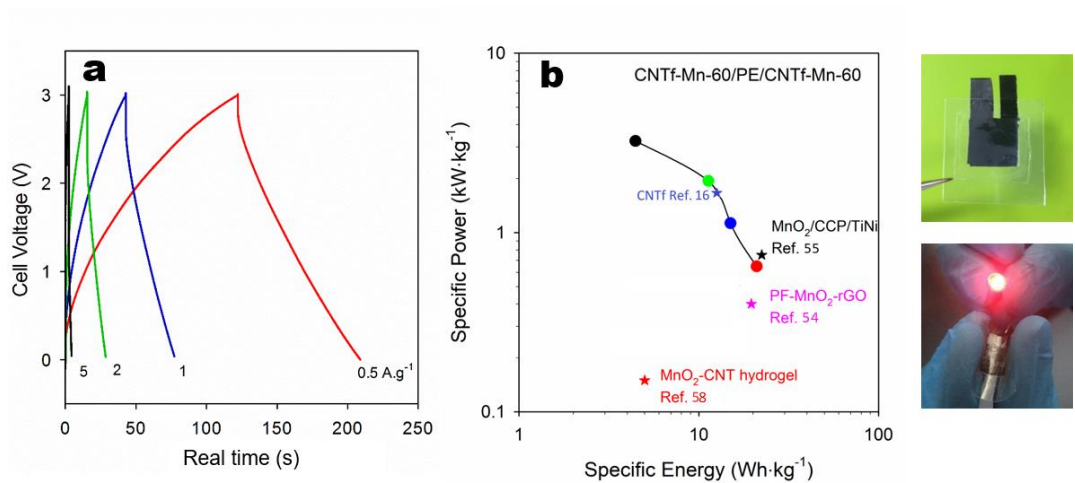

**Figure 8.** All-solid-state supercapacitors based on CNTf-MnO$_2$-60 symmetric electrodes and PYR$_{14}$TFSI-PVDF-co-HFP polymer electrolyte: (a) galvanostatic charge-discharge profiles of CNTf-Mn-60//CNTf-MnO$_2$-60 symmetric supercapacitors at various current densities in 3.0 V; (b) Ragone plot of all-solid-state supercapacitors based on total mass at 3.0 V in comparison with some relevant solid-state supercapacitors recently reported in literature. Photos show laminated all-solid-state supercapacitors capable of lighting up a 5 mm-diameter red LED in bended mode, demonstrating flexibility of the supercapacitor.

All-solid-state flexible supercapacitors (SSSC) capable of working in high voltage window of 3.0 V were also fabricated by sandwiching two symmetric CNTf-Mn-60 electrodes with PYR$_{14}$TFSI based polymeric electrolyte (photographs shown in Figure 8). Figure 8a displays GCD profiles of the SSSC at various specific currents ranging from 0.5 to 5 A·g$^{-1}$. Accordingly, a high capacitance of 17 F·g$^{-1}$ (based on total mass of electrodes) was achieved at a specific current of 0.5 A·g$^{-1}$ (corresponding to 68 F·g$^{-1}$ for a single electrode). Ragone plot of the SSSC device has been depicted in Figure 8b and compared with some recently reported solid state supercapacitors, showing a maximum specific energy of 21 Wh·kg$^{-1}$ (real energy density of 15.6 Wh·kg$^{-1}$) and a maximum specific power of 8 kW·kg$^{-1}$ (real power density of 3.2 kW·kg$^{-1}$) at 3.0 V. Recently, the volumetric energy ($E_v$) and power density ($P_v$) based on the whole supercapacitor weight/volume (including electrodes, electrolyte, current collectors,) was recommended as more realistic parameter to assess the real potential of an active material [52]. Hence, we also calculated the real volumetric energy density and real volumetric power density of the full cell (including CNTf, MnO$_2$ deposit, and the polymer membrane), resulting in a high value of 1.7 mWh·cm$^{-3}$ (max energy of 2.1 mWh·cm$^{-3}$) and 245 mW·cm$^{-3}$ (max power of 430 mW·cm$^{-3}$), respectively. Table 1 summarizes the energy and power characteristics of recent flexible solid state supercapacitors based on MnO$_2$ as active electrode material. As it is seen, the achieved energy and power with our CNTf-MnO$_2$//CNTf-MnO$_2$ solid state supercapacitors are superior or comparable (but yet much easier fabricated in comparison with multi-step preparation route used in reference [54] or [55] in which a metallic alloy has been used as the current collector). The excellent performance of CNTf-MnO$_2$//CNTf-MnO$_2$ solid state supercapacitor is attributed to high voltage window of 3.0 V (the highest voltage window reported for MnO$_2$-based SCs), the absence of current collector, as well as excellent electrochemical properties of decorated CNTf samples with MnO$_2$ NPs. The high volumetric energy and power densities can be also due to well distributed porous $MnO_2$ nanoparticles all along the thickness of CNT fibers which assures high mass ratio of $MnO_2$/carbon network and full utilization of pseudocapacitance of the metal oxide. It can be also seen in Figure 8 that a single symmetric $CNTf-MnO_2$//$CNTf-MnO_2$ solid state supercapacitor could successfully power a red 5-mm diameter LED (20 mA) even in the bent state which clearly confirms viability of the full device under tension. All these results demonstrate promise of decorated $MnO_2$ CNT fibers as high performance electrode materials in various media from aqueous to IL-based solid state membrane.

**Table 1.** Performance comparison of $CNTf-MnO_2$ symmetric all-solid-state supercapacitor (SSSC) with recent previous reports on flexible SSSC with planar shape.

| Cell Structure | E (V) | Electrolyte | $E_{max}^{\alpha}$ (Wh·kg$^{-1}$) | $P_{max}^{\alpha}$ (kW·kg$^{-1}$) | $E_v$ (mWh·cm$^{-3}$) | $P_v$ (mW·cm$^{-3}$) |
|---|---|---|---|---|---|---|
| $CNTf-MnO_2$//$CNTf-MnO_2$ | 3.0 | $PYR_{14}TFSI$-PE | 21 | 8 | 2.1 | 430 |
| $Ni/MnO_2$//Ni/AC [13] | 2.5 | PVA-$Na_2SO_4$ | - | - | 0.78 | 2.5 |
| $CC^{\dagger}-MnO_2$//$CC-CoSe_2$ [53] | 1.6 | PVA-LiCl | - | - | 0.58 | 282 |
| $PF^{\varepsilon}-MnO_2$-rGO//$PF-MnO_2$-rGO [54] | 0.8 | PVA-$Na_2SO_4$ | 19.6 | 2.4 | - | - |
| $MnO_2$//$Fe_2O_3$-PPy [14] | 1.6 | PVA-LiCl | - | - | 0.22 | 165 |
| $MnO_2/CCP^{\xi}/TiNi$// $MnO_2/CCP/TiNi$ [55] | 2.5 | (PVDF-HFP)-[EMIM]$BF_4$ | 22.4 | 7.5 | - | - |
| $NG^{\S}-MnO_2$//AC [56] | 1.8 | PVA-LiCl | - | - | 3.5 | 180 |
| $MnO_2$-AC//AC [57] | 2.0 | PVA-LiCl | - | - | 0.766 | 143 |
| $MnO_2$/hydrogel like CNT film [58] | 1.0 | $Na_2SO_4$ | 5 | 3.5 | - | - |

$^{\alpha}$ $E_{max}$ and $P_{max}$ were calculated based on $E_{max}=1/2CV^2$, and $P_{max}=E_{max}/t_{discharge}$, respectively in order to compare with previous reported results; †CC = Carbon Cloth; ‡ AC = Activated Carbon; ε PF = Paper-like Fiber; ξ CCP = Conductive Carbon Paste; * § NG = Nitrogen-doped Graphene.

## 4. Conclusions

In conclusion, dandelion-shaped $MnO_2$ nanoparticles were uniformly decorated on macroscopic CNT fibers via a simple electrodeposition technique, forming free-standing high $MnO_2$/carbon mass ratio electrodes. Electrochemical properties of CNTf-$MnO_2$ electrodes as supercapacitor materials were tuned in neutral aqueous electrolyte, achieving a high capacitance of 87 $F·g^{-1}$ per total mass of electrode. Moreover, assembling symmetric supercapacitors based on CNTf-$MnO_2$ electrodes with pure $PYR_{14}TFSI$ ionic liquid resulted in high performance supercapacitors working at 3.0 V with high real specific energy of 36 $Wh·kg^{-1}_{MnO_2}$ and real specific power of 17 $kW·kg^{-1}_{MnO_2}$, possessing a remarkable cycling stability without any significant capacitance fade over long cycling at 2 $A.g^{-1}$. Finally, flexible all-solid-state supercapacitors were assembled with using $PYR_{14}TFSI$-based polymer electrolyte, capable of operating at 3.0 V and offering remarkable energy density of 21 $Wh·kg^{-1}$ (2.1 $mWh·cm^{-3}$) and power density of 8 $kW·kg^{-1}$ (430 $mW·cm^{-3}$). Thus, this study demonstrates an effective and facile method in the fabrication of free-standing high dense $MnO_2$/carbon nanotube fiber lightweight films as electrode of energy storage devices for practical applications.

**Notes**

Supporting Information is available online or from the author.

**ACKNOWLEDGMENT**

Authors gratefully acknowledge financial support from MINECO (former MICINN) through the MAT2015-64167-C2-1-R project, European Union structural funds and the Comunidad de Madrid MAD2D-CM Program (S2013/MIT-3007). JJV is grateful for generous financial support provided by the European Union Seventh Framework Program under grant agreements 678565 (ERC-STEM) and by MINECO (RyC-2014-15115).